\documentclass[5p]{elsarticle}

\usepackage{hyperref}
\usepackage{amsmath}
\usepackage{amssymb}
\usepackage{xfrac}
\usepackage{wasysym}
\usepackage{multirow}
\usepackage{booktabs}

\usepackage[table,xcdraw]{xcolor}

\journal{Nuclear Instruments and Methods in Physics Research Section A}

\begin{document}

\begin{frontmatter}

\title{Plastic Scintillation Detectors for Time-of-Flight Mass Measurements}

\author[CMU,IMPCAS,JINA]{K. Wang\corref{corresponding}}\ead{wang5k@cmich.edu}
\author[CMU,JINA,NSCL]{A. Estrade\corref{corresponding}}\ead{estra1a@cmich.edu}
\author[CMU]{S. Neupane\fnref{UTK}}
\author[CMU]{M. Barber}
\author[WMU,JINA]{M. Famiano}
\author[NSCL]{T. Ginter}
\author[CMU]{D. McClain\fnref{TAMU}}
\author[CMU,JINA]{N. Nepal}
\author[NSCL]{J. Pereira}
\author[NSCL,JINA]{H. Schatz}
\author[CMU]{G. Zimba\fnref{JYU}}

\cortext[corresponding]{Corresponding authors}
\fntext[UTK]{Present address: Department of Physics and Astronomy, University of Tennessee, Knoxville, Tennessee 37996, USA}
\fntext[TAMU]{Present address: Department of Physics and Astronomy, Texas A\&M University, College Station, Texas 77843, USA}
\fntext[JYU]{Present address: Department of Physics, University of Jyv\"{a}skyl\"{a}, P.O. Box 35, FI-40014,  Jyv\"{a}skyl\"{a}, Finland}

\address[CMU]{Department of Physics, Central Michigan University, Mount Pleasant, Michigan 48859, USA}
\address[IMPCAS]{Institute of Modern Physics, Chinese
Academy of Sciences, Lanzhou, Gausu 730000, China}
\address[JINA]{Joint Institute for Nuclear Astrophysics Center for the Evolution of the Elements (JINA-CEE), USA}
\address[NSCL]{National Superconducting Cyclotron Laboratory, Michigan State University, East Lansing, Michigan 48824, USA}
\address[WMU]{Department of Physics, Western Michigan University, Kalamazoo, Michigan 49008, USA}

\begin{abstract}
Fast timing detectors are an essential element in the experimental setup for time-of-flight (ToF) mass measurements of unstable nuclei.
We have upgraded the scintillator detectors used in experiments at the National Superconducting Cyclotron Laboratory (NSCL) by increasing the number of photomultiplier tubes that read out their light signals to four per detector, and characterized them in a test experiment with $^{48}$Ca beam at the NSCL.
The new detectors achieved a time resolution ($\sigma$) of 7.5 ps.
We systematically investigated different factors that affect their timing performance.
In addition, we evaluated the ability of positioning the hitting points on the scintillator using the timing information and obtained a resolution ($\sigma$) below 1 mm for well-defined beam spots. 
\end{abstract}

\begin{keyword}
Time-of-flight mass measurement; Plastic scintillator; Photomultiplier tubes; Time-walk correction
\end{keyword}

\end{frontmatter}

\section{Introduction}

Nuclear masses, and nuclear binding energies, play a central role in many questions of nuclear structure and nuclear astrophysics \cite{Lunney2003,Famiano2019}.
Nuclear masses provide one of the main tools to understand the evolution of nuclear structure away from $\beta$-stability through systematic trends in binding energies \cite{Zhang1989,Brown2013}, and are an essential input for nuclear astrophysics models \cite{Mumpower2016,Horowitz2019}.

At present there is a variety of techniques and devices capable of measuring the mass of isotopes at different regions across the nuclear chart, and with various degrees of precision: Penning trap spectrometers \cite{Mukherjee2008,Block2007,Eronen2012,Ketelaer2008,Savard2006,Redshaw2013,Dilling2006}, storage rings \cite{Zhang2016,Bosch2013,Xu2013,Yamaguchi2015}, multi-reflection time-of-flight (MR-ToF) spectrometers \cite{Wolf2013,Schury2013,Dickel2015,Hirsh2016,Jesch2015}, and time-of-flight measurements with magnetic spectrometers (ToF-$B\rho$ technique) \cite{Matos2012,Meisel2013}.
The latter has a relatively low mass resolving power with $m/\Delta m\sim10^4$, but can measure with high efficiency the masses of many unstable isotopes far from $\beta$-stability.
The technique is currently used with the S800 spectrometer at the National Superconducting Cyclotron Laboratory (NSCL) \cite{Matos2012,Meisel2016}, which is the focus of this work, and with the SHARAQ spectrometer at RIKEN \cite{Michimasa2018}.

The principle of the ToF-$B\rho$ technique is based on the motion law of an ion with mass $m$, charge $q$ and momentum $p$ passing through a beam line and magnetic spectrometer with a total flight path of length $L$.
If its time-of-flight is given by $T$, the nuclear mass is related to these variables by:

\begin{equation} \label{mass}
	m=p\cdot\sqrt{\left(\frac{T}{L}\right)^2-\frac{1}{c^2}}=qB\rho\cdot\sqrt{\left(\frac{T}{L}\right)^2-\frac{1}{c^2}},
\end{equation} 
where $c$ is the speed of light, and $B\rho=p/q$ is the magnetic rigidity of this ion with radius of curvature $\rho$ for the particle trajectory.

In order to obtain masses, $T$ must be measured with very high precision using timing detectors at the start and end points of the flight path.
The momentum $p$ can be obtained from measuring the ion's position $x$ in a dispersive plane of the spectrometer.
To first order:

\begin{equation} \label{momentum}
	p=p_0\left(1+\frac{x}{D}\right),
\end{equation}
where $p_0=q(B\rho)_0$ is the momentum of the central trajectory, and $D$ is the dispersion function.
The electric charge $q$ of the beam ions is evaluated with a relation based on the total kinetic energy, velocity and magnetic
rigidity combining with the $\Delta E$-ToF particle-identification technique \cite{Tarasov2009}.

However, $L$ and $B\rho$ usually cannot be measured with sufficient accuracy. Therefore, in practice, we can derive the mass by expanding $m/q$ in Eq. \eqref{mass} as a polynomial function of the measured parameters ($T$, $x$, etc.), and then determine them from the information of known-mass reference nuclides.

\begin{figure}[htbp]\centering
	\includegraphics[width=0.9\columnwidth]{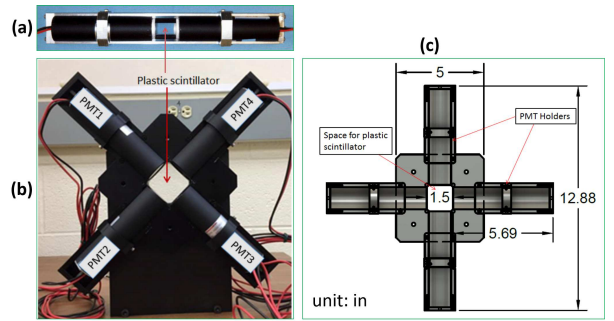}
	\caption{(color online). Photographs of (a) timing detector used in previous ToF-$B\rho$ experiments at the NSCL \cite{Matos2012} and (b) new timing detector studied in this work. The engineering drawing (c) for the new detector frame shows the dimensions of its design \cite{Neupane2017}.} \label{plastic}
\end{figure}

At the NSCL, the experiments use the S800 spectrograph operated in dispersion-matched mode.
The ToF from the final focal plane of the A1900 fragment separator to the S800 focal plane is measured with two fast-timing plastic scintillation detectors.
In the previous experiments \cite{Estrade2011,Matos2012,Meisel2015,Meisel2015a,Meisel2016}, each timing detector consists of one thin fast organic scintillator with two photomultiplier tubes (PMTs) coupled to its opposite sides, as shown in Fig. \ref{plastic} (a).
The $B\rho$ at the dispersive plane at the target position of the S800 spectrograph is measured with a micro-channel plate (MCP) detector.
The energy loss $\Delta E$ is measured by a ionization chamber or silicon detectors at the final focal plane \cite{Meisel2016}.
Here we present the development of an upgraded system of timing detectors. Details of the MCP detector can be found in Refs. \cite{Meisel2016,Rogers2015}.

The main contributions to the mass resolution include time resolution of timing detectors and position resolution of the MCP detector.
The time resolution $\sigma_T$ of the previous timing detectors was measured to be $\sim$30 ps with primary beam tests \cite{Estrade2010,Matos2012}.
For the typical flight time of $\sim$500 ns in the S800 experiments the contribution of time resolution to the final mass resolution is $\frac{\left(\sfrac{\sigma_T}{T}\right)}{1-\left(\sfrac{L}{cT}\right)^2}\approx7\times10^{-5}$.
In addition, the currently achieved position resolution of $\sim$0.5 mm of the MCP detector at the dispersive plane with the dispersion function of $\sim$11 cm/\% results in a momentum resolution of $\sfrac{\sigma_p}{p}\approx5\times10^{-5}$ according to Eq. \eqref{momentum} \cite{Matos2012,Meisel2016}, which contributes to the final mass resolution as $5\times10^{-5}$.
Besides the resolution of the beamline detectors, other factors like beam straggling in the detectors and variations in the flight path followed by each beam particle also affect the mass resolution.

From the above analysis, we can see that one main contribution affecting the mass resolution is the time resolution.
In order to approach the realm of mass resolution of $10^{-5}$, an important step is to improve the timing performance of the timing detector.
In fact, similar detectors with an intrinsic $\sigma_T$ of about $\sim$10 ps have been developed and tested with ion beams \cite{Nishimura2003,Zhao2016}.
Large plastic scintillators read out by many PMTs have also been successfully tested and achieved picosecond resolution \cite{Hoischen2011,Ebran2013}.
Silicon photomultipliers (SiPMs) provide an alternative to PMTs to the scintillator signal and have been shown to achieve comparable timing resolution \cite{Cattaneo2016,Bohm2018,Cortes2018}.
While their main advantages of compact design and small power requirements are not decisive factors in our application, it would be interesting to explore their use for ToF mass measurement applications in the future.
Our work can serve as a comparison benchmark for such developments.
We note that before introducing new detector systems care has to be taken to minimize the systematic errors at the picosecond level, which could affect the mass measurement results.

These results motivate our design for a new detector where each plastic scintillator is coupled to 4 PMTs as shown in Fig. \ref{plastic} (b).
We expect that doubling the number of PMTs  will result in a significant increase in the number of photoelectrons produced in the PMTs ($N_{\text{p.e.}}$), which is an important limiting factor for the resolution because our detectors use thin scintillators to minimize beam straggling.
An improvement of $\sigma_T$ by a factor of $\sfrac{1}{\sqrt{2}}\sim0.7$ would be expected from the relationship $\sigma_T\propto\sfrac{1}{\sqrt{N_\text{p.e.}}}$ discussed in \cite{Ebran2013}.

\section{Detector design}

Each timing detector consists of four photomultiplier tubes (PMTs) attached to a thin organic scintillator with a surface area of $4\times4$ cm$^2$, as shown in Fig. \ref{plastic} (c).
This design doubles the number of PMTs and quadruples the surface area compared to the detectors previously used for ToF-$B\rho$ experiments at the NSCL.

As the choice of scintillator material, we used BC-418 produced by Saint-Gobain \cite{SaintGobain}, which was used by previous detectors.
Before testing the detectors with a fast ion beam, the new design was characterized offline using a table-top laser setup at Central Michigan University (CMU).
Details about it can be found in Neupane’s thesis \cite{Neupane2017}.
During this test we also studied the EJ-228 and EJ-232 scintillators from Eljen Technology \cite{EljenTechnology}.
The properties of these scintillators are shown in Table 1.
Their main differences are the higher light output of BC-418 and EJ-228, and the faster rise time of the signal in EJ-232 (both of which are desirable properties for fast timing).
The laser test showed similar time resolution for the three scintillators, from 6.7 ps to 6.3 ps, showing that the choice among them is not a dominant contribution to the resolution of our design.
To decrease the time spread of photons transmitted from the scintillator to the PMT, a small size is preferable.
In addition, the thickness of the scintillators was chosen to be 0.5 mm to avoid large beam straggling during mass measurement experiments.

For PMTs, the R4998 type integrated in the H6533 assembly provides adequate capabilities as shown in Table \ref{tab2}, including large gain and short signal rise time \cite{Hamamatsu}.
The PMTs are coupled to the scintillators with BC-634A silicon pads from Saint-Gobain \cite{SaintGobainPad} ($\diameter$25.4 mm $\times$ 3 mm), which have similar optical properties as the scintillator and the photocathode of the PMT to provide good coupling between the scintillator's edge and the PMT.

An assembled detector is shown in Fig. \ref{plastic} (b).

\begin{table}[htbp]
    \centering
    \caption{Characteristics of plastic scintillators BC-418 from Saint-Gobain \cite{SaintGobain}, and EJ-228 and EJ-232 from Eljen Technology \cite{EljenTechnology}.}
    \label{tab1}
    \begin{tabular}{@{}cccc@{}}
    \toprule
    Scintillator         & BC-418     & EJ-228     & EJ-232 \\ \midrule
    \begin{tabular}[c]{@{}c@{}}Light Output\\ (\%Anthracene)\end{tabular}      & \multicolumn{2}{c}{67} & 55   \\
    \begin{tabular}[c]{@{}c@{}}Efficiency\\ (photons/1 MeV e$^-$)\end{tabular} & ---       & 10200      & 8400 \\
    Rise Time (ns)       & \multicolumn{2}{c}{0.5} & 0.35   \\
    Decay Time (ns)      & \multicolumn{2}{c}{1.4} & 1.6    \\
    Pulse Width (ns)     & \multicolumn{2}{c}{1.2} & 1.3    \\
    Max. Wavelength (nm) & \multicolumn{2}{c}{391} & 370    \\
    Density (g/cm$^3$)   & \multicolumn{3}{c}{1.023}        \\
    Refractive Index     & \multicolumn{3}{c}{1.58}         \\ \bottomrule
    \end{tabular}
\end{table}

\begin{table}[htb]\centering
    \caption{Characteristics of R4998 PMT within H6533 assembly from Hamamatsu \cite{Hamamatsu}.}\label{tab2}
    \begin{tabular}{@{}cc@{}}
    \toprule
        Property & Value \\ \midrule
        Assembly Size & $\diameter$31 mm \\
        PMT Tube Size & $\diameter$25 mm \\
        Anode-to-Cathode Voltage & -2250 V \\
        Wavelength Range & 300--650 nm \\
        Wavelength Peak & 420 nm \\
        Luminous Sensitivity & 70 $\mu$A/lm \\
        Quantum Efficiency & 13\%--39\% \\
        Gain & $5.7\times10^6$ \\
        Rise Time & 0.7 ns \\
        Transit Time & 10 ns \\
        Transit Time Spread & 0.16 ns \\ \bottomrule
    \end{tabular}
\end{table}

\section{Experiment}

After the offline tests using laser at CMU \cite{Neupane2017}, here we present results from a test performed with a fast beam of stable isotopes at the NSCL, which provided similar conditions to those of ToF-$B\rho$ mass measurement experiments.

\subsection{Detectors setup}

The experiment was performed in the S2 vault at NSCL \cite{Gade2016}.
A primary beam of $^{48}$Ca with an energy of 140 MeV/u passed through a beryllium target with a thickness of 1081 mg/cm$^2$, resulting in a  $^{48}$Ca beam with degraded energy of 90 MeV/u and 99 \% purity.
This beam energy was chosen for the energy loss in the scintillators to be similar to a planned experiment with isotopes of $Z\sim40$.
Straggling in the target increased the angular emittance of the beam, which helped to steer the beam to illuminate different spots in the scintillator surface.
The $^{48}$Ca beam was delivered to the detection station at rates around 500 particles per second during most time of the test.

\begin{figure}[htbp]\centering
	\includegraphics[width=0.9\columnwidth]{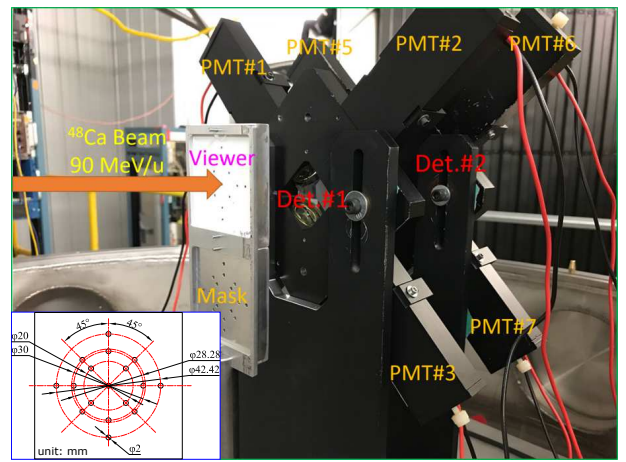}
	\caption{(color online). The detection setup inside the 53-inch chamber in S2 vault during the beam test. The inset shows the pattern of holes on the mask.}\label{setup}
\end{figure}

The detectors setup is shown in Fig. \ref{setup}.
The distance between two timing detectors was 9 cm.
An aluminum mask with a size of $6\times6\times1$ cm$^3$ was mounted $\sim$8 cm in front of the first timing detector.
The holes in the mask had a diameter of 2 mm with a pattern shown in the inset of Fig. \ref{setup}.
The mask performed two functions: (1) guiding the focused beam to interact with a small area of the scintillators when passing through one particular hole; (2) helping to obtain a position distribution for defocused-beam settings when the ion beam could pass through several holes.
In addition, a ZnS viewer was installed above the mask in a retractable ladder to tune the beam's positions for different settings.

\subsection{Electronics setup}

\begin{figure}[htbp]\centering
	\includegraphics[width=1.0\columnwidth]{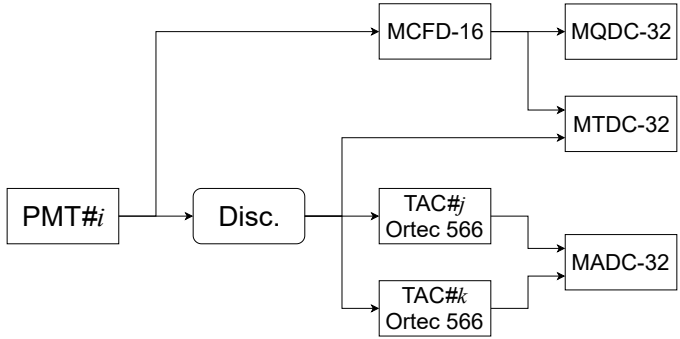}
	\caption{The schematic diagram of electronics used in the experiment. Totally, there are 8 PMTs marked by $i$ and 8 TACs marked as $j$ and $k$. The start and stop inputs of each TAC are signals from PMTs of the first ($i=1,2,3,4$) and second ($i=5,6,7,8$) timing detector. For these 8 TACs, the start-stop PMTs combinations are: (1) \#1---\#5, (2) \#2---\#6, (3) \#3---\#7, (4) \#4---\#8, (5) \#1---\#6, (6) \#2---\#7, (7) \#3---\#8, (8) \#4---\#5.}\label{electronics}
\end{figure}

The schematic diagram of electronics is shown in Fig. \ref{electronics}.
Signals from each PMT were split into two.
One signal was delivered to a 16-channel constant-fraction discriminator (CFD), the MCFD-16 module from Mesytec.
The modules provided two outputs for each channel: the analogue outputs were connected to a charge-to-digital converter (QDC) (MQDC-32 from Mesytec \cite{Mesytec}) to measure the amplitude of the PMT signals, while the timing discrimination outputs were connected to a time-to-digital converter (TDC) (MTDC-32 from Mesytec \cite{Mesytec}) to give the time information.
The other split PMT signal was fed to a timing discriminator, with different types employed during the test: leading-edge discriminators (LEDs, Phillips Scientific 704 and 711 \cite{Phillips}) or constant-fraction discriminators (CFDs, Ortec 935 \cite{Ortec} and Tennelec 455 \cite{Tennelec}).
Their timing information was processed by a MTDC-32 module, and also digitized by time-to-amplitude converters (TACs) connected to an analog-to-digital converter (ADC).
We used 8 TAC-ADC channels to measure relative time between different combinations of PMTs in the first and second detector, providing redundant information for the time-of-flight and allowing us to calculate the time difference between any two PMTs to derive position information (See Sec. \ref{posdist}). The modules of TACs and ADC were Ortec 566 \cite{Ortec} and Mesytec MADC-32 \cite{Mesytec}, respectively.
The trigger and gate for the data acquisition system were the OR logic of all PMT signals supplied by the MCFD-16.
Various timing modules and techniques regarding the electronics were systematically compared to obtain the setup with best timing performance. 

\section{Data analysis and results}

The goal of the test was to demonstrate the improved resolution of the new timing detector design, and investigate its dependence on the parameters such as high voltages of PMTs, type of discriminators and beam intensity. 
In addition, the defocused beam was used to investigate the position resolution of the plastic detectors.

\subsection{Time-of-flight analysis}\label{ToFAna}

The target variable of our measurement was the time-of-flight (ToF) of the beam ions across the 9-cm path between the first and second scintillator detectors.
Each detector provided four signals from its PMTs, from which we obtained different time measurements with the digitizer modules in our electronics setup (Figure \ref{electronics}).
The measured times can be combined in different ways to obtain a value for ToF.

For the signal processing with TAC+ADC combinations, the ToF is defined as the average of four independent combinations:

\begin{equation} \label{ToFTA}
	ToF_{Ta}=\frac{\sum\limits_{i=1}^{4}Ta_i}{4},
\end{equation}
where symbol \emph{Ta} denotes the time values from TAC+ADC setup (see Fig. \ref{electronics} for the corresponding channel assignments).
We can also obtain a measurement of the time-of-flight when averaging the values of $Ta_5$ to $Ta_8$ of the other four TAC+ADC channels.

For the signal processing with TDC, the timing signals of start and stop timing detectors are taken as the average time from four PMTs of each scintillator and then ToF is defined as:

\begin{equation} \label{ToFTDC}
	ToF_{Tt}=\frac{\sum\limits_{i=5}^{8}Tt_i}{4}-\frac{\sum\limits_{i=1}^{4}Tt_i}{4},
\end{equation}
where symbol \emph{Tt} denotes the timing signals from TDC setup.
The reason for averaging all the timing signals of PMTs coupled to one scintillator in Eqs. \eqref{ToFTA} and \eqref{ToFTDC} is to minimize the timing uncertainty introduced by different hitting positions of beam on the plastic, and to obtain a better resolution.

In order to correct the time-walk effect due to the variance in signal amplitudes of PMTs, especially for the leading-edge discriminator (LED) timing method, we use the following correcting equation:

\begin{equation} \label{ToFCorr}
	ToF_\text{corr}=ToF_\text{raw}+\left[ToF_\text{pivot}-f_{ToF}(\mathbf{Q})\right],
\end{equation}
where $ToF_\text{raw}$ is the time measured with the TAC+ADC ($ToF_{Ta}$) or TDC ($ToF_{Tt}$) setup.
$ToF_\text{pivot}$ represents the pivot point of ToF to be realized in the ideal case without time walk.
$f_{ToF}(\mathbf{Q})$ stands for the dependence of the measured ToF on the integrated charge ($\mathbf{Q}$) of each PMT signal recorded by the QDC.
Note that here $\mathbf{Q}$ represents the information of signal amplitudes for all eight PMT channels.

We find that a linear function for $f_{ToF}(\mathbf{Q})$ provides a good fit:

\begin{equation} \label{fToF}
	f_{ToF}(\mathbf{Q})=c_0+\sum\limits_{i=1}^{8}c_i\cdot Q_i,
\end{equation}
Here $c_0$ and $c_i$ are the correction parameters to be determined by fitting the correlation between $ToF_\text{raw}$ and the charge of each PMT, $Q_i$.
Then, $ToF_\text{pivot}$ is taken as $c_0+\sum\limits_{i=1}^{8}c_i\cdot\overline{Q_i}$, where $\overline{Q_i}$ means the average charge of $i^\text{th}$ PMT for all events under the same condition.

It is worth noting that the choice of $ToF_\text{pivot}$ only leads to a global systematic shift of the centroid of raw ToF distribution, without impacting the time resolution. Therefore, comparisons of the measured ToF centroids are only meaningful if the time-walk correction was done with the same set of parameters (i. e. same choice of $ToF_\text{pivot}$).
Thus, Eq. \eqref{ToFCorr} can be rewritten as:

\begin{equation} \label{ToFCorr2}
	ToF_\text{corr}=ToF_\text{raw}+\left[\sum\limits_{i=1}^{8}c_i\cdot \overline{Q_i}-\sum\limits_{i=1}^{8}c_i\cdot Q_i\right].
\end{equation}

In order to more clearly show the time-walk effect, we introduce the charge difference $\Delta Q$ between PMTs of each scintillator as:

\begin{equation} \label{DeltaQ}
	\Delta Q=\frac{\sum\limits_{i=5}^{8}Q_i}{4}-\frac{\sum\limits_{i=1}^{4}Q_i}{4}.
\end{equation}
Then the relationship between ToF and $\Delta Q$, instead of the charge of single PMT, provides a better visualization of the magnitude of the time-walk effect.

\begin{figure}[htbp]\centering
	\includegraphics[width=0.9\columnwidth]{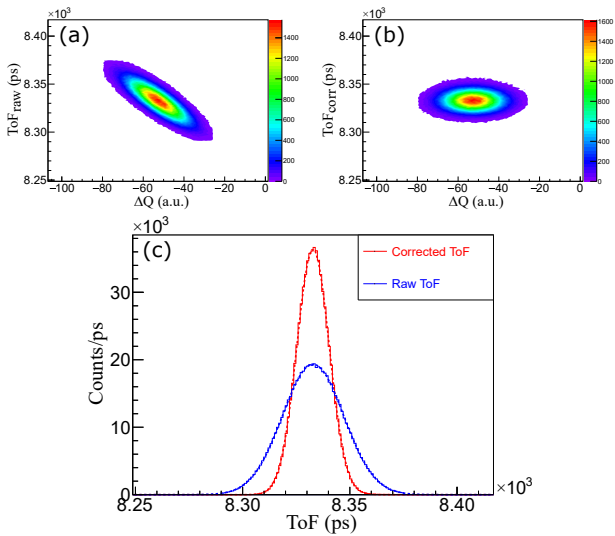}
	\caption{(color online). Correlation between ToF obtained via LED+TAC+ADC setup and the amplitude difference $\Delta Q$ (a) without and (b) with time-walk correction. Panel (c) shows the distributions of raw (blue) and corrected (red) ToFs fitted with Gaussian functions indicated by the corresponding dash lines. The data corresponds to setting (2) in Table \ref{tab3}.} \label{TofComp}
\end{figure}

As an example, Fig. \ref{TofComp} (a) shows that there is an obvious dependence of ToF on $\Delta Q$, calculated with the time values measured with the leading-edge discriminator combining with TAC and ADC electronic modules.
After using the above correction method, the time-walk effect of ToF on the signal amplitude is removed as shown in Fig. \ref{TofComp} (b).
By comparing the ToF distributions before and after correction in Fig. \ref{TofComp} (c), we can see there is a clear improvement in the time resolution.
It is improved from 14.2 ps to 7.5 ps (about 50\%) for the LED+TAC+ADC electronics setup of Fig. \ref{TofComp}.
Note that throughout this paper we measure the resolution as the sigma of a Gaussian fit to the $ToF$ distribution.

\begin{table*}[htbp]\centering
    \caption{ToF properties at different conditions in the present experiment.$^a$}
    \label{tab3}
    \begin{tabular}{@{}cccccccc@{}}
    \toprule
    \multirow{2}{*}{Setting} &
      \multirow{2}{*}{PMT HV$\downarrow\uparrow^b$} &
      \multirow{2}{*}{Electronics} &
      \multirow{2}{*}{\begin{tabular}[c]{@{}c@{}}Beam\\ Position [mm, mm]\end{tabular}} &
      \multicolumn{2}{c}{ToF Centroids [ps]$^c$} &
      \multicolumn{2}{c}{\begin{tabular}[c]{@{}c@{}}Time Resolution\\ ($\sigma$) [ps]\end{tabular}} \\ \cmidrule(l){5-8} 
         &                  &             &                      & Raw   & Corrected & Raw  & Corrected \\ \midrule
    (1)  & 0                & LED+TDC     & (0, 0)               & 0     & 0.0       & 16.6 & 11.7      \\
    (2)  & 0                & LED+TAC+ADC & (0, 0)               & 0     & 0.0       & 14.2 & 7.5       \\
    (3)  & 0                & LED+TAC+ADC & (-7, 7)              & 10.8  & 0.3       & 14.1 & 8.6       \\
    (4)  & 0                & LED+TAC+ADC & (14, 0)              & -25.4 & 2.6       & 14.3 & 8.0       \\
    (5)  & 0                & LED+TAC+ADC & (21, 0)              & 21.9  & 2.4       & 20.8 & 9.5       \\
    (6)  & 0                & LED+TAC+ADC & Defocused$^d$        & ---  & ---      & --- & 8.6       \\
    (7)  & 0                & CFD+TAC+ADC & (0, 0)               & 0     & 0.1       & 11.2 & 10.3      \\
    (8)  & 0                & CFD+TAC+ADC & (-7, 7)              & -5.4  & -0.6      & 12.2 & 11.3      \\
    (9)  & 0                & CFD+TAC+ADC & (14, 0)              & -0.4  & -0.3      & 12.6 & 10.2      \\
    (10) & 0                & CFD+TAC+ADC & Defocused            & ---  & ---      & --- & 12.2      \\
    (11) & $\downarrow$50 V & CFD+TAC+ADC & (0, 0)               & 9.1   & ---       & 11.6 & 10.7      \\
    (12) & $\uparrow$50 V   & CFD+TAC+ADC & (0, 0)               & 23.0  & ---      & 11.0 & 10.2      \\
    (13) & $\uparrow$50 V   & CFD+TAC+ADC & (0, 0) with 10 kHz   & 21.5  & ---      & 11.6 & 10.6      \\
    (14) & $\uparrow$50 V   & CFD+TAC+ADC & (0, 0) with >100 kHz & ---   & ---       & 299  & 299       \\ \bottomrule
    \end{tabular}
    \flushleft
	$^{a}$ Errors of all time values are less than 0.1 ps. \\
	$^{b}$ The changes of high voltages supplied to PMTs are relative to the values of setting (1). \\
	$^{c}$ ToF raw values are shifted relative to the centroid of raw ToF distribution using the same electronics and beam at position (0, 0). Values for the corrected ToF centroids are only given for settings where the same walk correction parameters are applied (settings 2 to 5, and 7 to 9).  \\
	$^{d}$ In the defocused setting the beam covered at least four left holes at (-7, $\pm$7) and (-11, $\pm$11) through the mask. No ToF centroids are reported since the spectrum is a mixture of distributions from beam ions through different mask holes.\\
\end{table*}

The summary of data analysis results are displayed in Table \ref{tab3}.
In the following subsections, from \ref{comp1} to \ref{comp3}, we present the results from measurement settings with a focused beam impinging on the center of timing detectors.
Subsections \ref{comp4} and \ref{posdist} present the discussions to study position dependent effects in the response of the detectors, and their position resolution.
In subsections \ref{comp5} and \ref{comp6} we present the response to bias voltage and beam rate.

\subsection{Comparison between two- and four-PMT readouts}\label{comp1}

\begin{figure}[htbp]\centering
	\includegraphics[width=0.9\columnwidth]{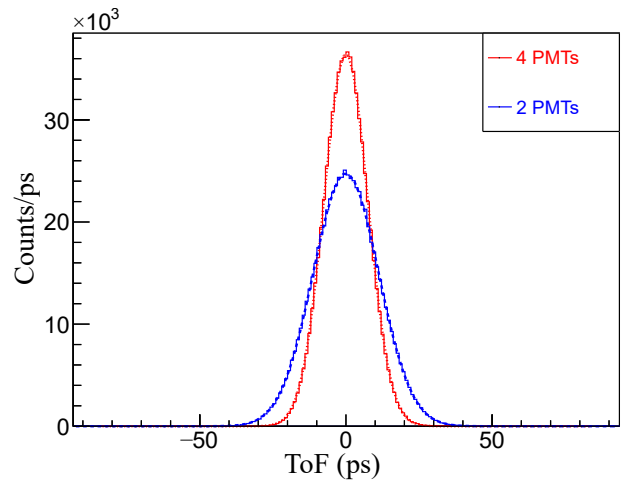}
	\caption{(color online). Comparison between the distributions of time-walk corrected ToFs obtained from 4 PMTs (red) and 2 PMTs (blue) using the same LED+TAC+ADC electronics setup. The distributions are fitted with Gaussian functions plotted with the responding colors. For better comparing display, both distributions are shifted to have a mean ToF of zero.}\label{TofComp4_2}
\end{figure}

We begin the discussion of the ToF resolution results by comparing the value obtained by using all four PMTs in each detector with that using two of the PMTs, which is comparable to the detector design used in previous mass measurement experiments.
The time-walk correction for the two-PMT ToF is similar to that for the four-PMT ToF except for using time and amplitude information of two PMTs of each detector. 
The comparison between these two ToF distributions with the LED+TAC+ADC electronics setup is illustrated in Fig. \ref{TofComp4_2}, which clearly indicates the time resolution taken from 4 PMTs is better than that from 2 PMTs in agreement with our expectation.
The time resolution with 4-PMT readout should be improved to $\sfrac{1}{\sqrt{2}}\sim0.7$ times of that with 2-PMT readout \cite{Ebran2013}, which is well supported by the values we obtain here (from 11.2 ps to 7.5 ps).

\subsection{Comparison between TAC+ADC and TDC}

\begin{figure}[!ht]\centering
	\includegraphics[width=0.9\columnwidth]{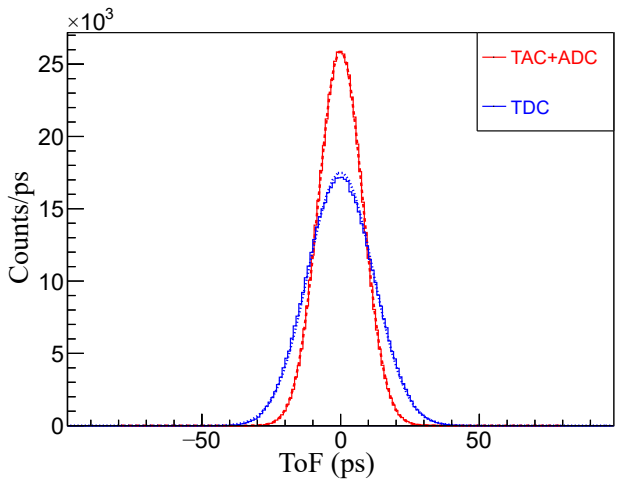}
	\caption{(color online). Comparison between the distributions of time-walk corrected ToFs recorded from TAC+ADC (red) and TDC (blue) using the same leading-edge discriminator. The Gaussian fitting functions are also plotted with the responding colors. For better comparing display, both ToF distributions are shifted to the position centered at 0. The data corresponds to settings 1 and 2 in Table \ref{tab3}.}\label{TofCompA_T}
\end{figure}

A time-to-digital converter (TDC) was also used during the experiment.
A TDC can simplify electronics set up since a single module can replace many TACs and one ADC.

We compare the effects on the timing performances of these two electronic setups used to digitize the time measurement.
Fig. \ref{TofCompA_T} shows the distributions of ToFs obtained from TAC+ADC and TDC after time-walk correction.
It is found that the time resolution with TAC+ADC (7.5 ps, setting (2) in Table \ref{tab3}) is $\sim$35\% better than that with TDC (11.7 ps, setting (1) in Table \ref{tab3}).

The reason for this can be related to the difference of the time measurement techniques used in these two electronics modules.
For the TAC+ADC, the TAC uses a high-precision analog technique to convert the time interval to pulse amplitude, which is then converted to a digital signal by the ADC.
For the TDC, the same analog technique is only used for the time interval smaller than the period of clock counter inside TDC, which measures the longer intervals.
In fact, according to the manuals of TAC (ORTEC 566 \cite{ORTEC-566}) and TDC (Mesytec MTDC-32 \cite{MTDC-32}), the time resolution ($\sigma$) of the former ($\leqslant$4.3 ps for 50 ns range) is better than that of the latter (5--10 ps).
Therefore, although the TDC leads to a simpler setup, the TAC+ADC combination provides the better time resolution.

\subsection{Comparison between LED and CFD timing techniques}\label{comp3}

\begin{figure}[htbp]\centering
	\includegraphics[width=0.9\columnwidth]{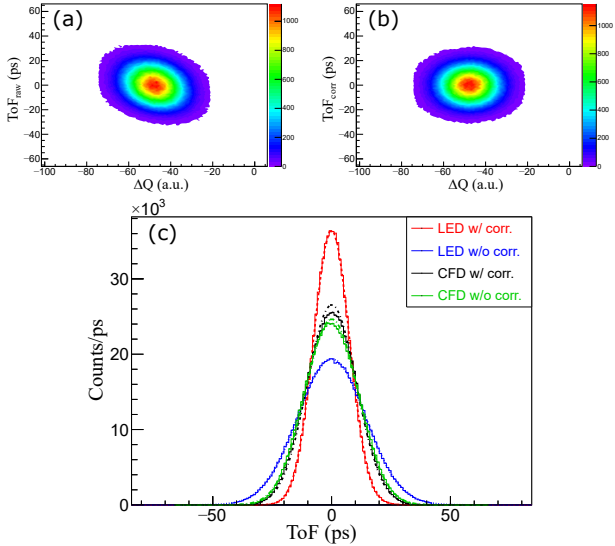}
	\caption{(color online). Correlation between ToF obtained by CFD+TAC+ADC electronics setup and the amplitude difference $\Delta Q$ (a) without and (b) with the time-walk correction. Panel (c) shows the comparison between the distributions of ToFs with LED (red and blue for corrected and raw distributions, respectively) and CFD (black and green for corrected and raw distributions, respectively) timing techniques. The Gaussian fitting functions are also plotted with the responding colors. For better comparing display, all the four ToF distributions are normalized to the same total counts and shifted to the position centered at 0. The data corresponds to settings 2 and 7 in Table \ref{tab3}.} \label{TofCompL_C}
\end{figure}

Constant-fraction discriminators (CFDs) address the time-walk issues and minimize the dependence of ToF on the signal amplitudes \cite{Gedcke1968}.
In our measurement we used Ortec 935 and Tennelec 455 CFDs with a 1 ns delay. Indeed, Fig. \ref{TofCompL_C} (a) shows a much smaller correlation between ToF and amplitude difference $\Delta Q$ when compared with the measurement using LEDs (Fig. \ref{TofComp}).
After the time-walk correction presented in Fig. \ref{TofCompL_C} (b) for the CFD data set, Fig. \ref{TofCompL_C} (c) shows that there is a relatively small improvement in the time resolution, from 11.2 ps to 10.3 ps (setting (7) in Table \ref{tab3}), when using CFDs instead of LEDs.
Moreover, time resolution from the CFD method without the correction is already close to 10 ps and better than that from LED method.

The $\sim$1 ps residual time walk when using CFD method may result from the charge sensitivity of the zero crossover comparator inside the CFD module \cite{Regis2012,Paulus1985}. In addition, it is difficult to adjust the settings of CFD with detector signals produced by the ion beam, since the experimental area is not accessible when the beam is on.
A pulse generator (Phillips Scientific 417 \cite{Phillips}) with signal rise time of $\sim$1.5 ns and amplitude of -800 mV was used to adjust the $\sim$1 ns total delay and walk setting for the CFDs.
It is possible that theses setting parameters are not the optimum for beam-induced signals from the PMTs, which leads to the remaining time walk of the CFD measurement.

\subsection{Comparison between different beam positions}\label{comp4}

Up to now, the discussion concentrated on the data obtained with the $^{48}$Ca beam focused at the center of the plastic scintillator, after passing through the central hole of the collimator mask. 
In a mass measurement experiment, however, the fragmentation beams typically have a larger cross-sectional area of $\sim$1 cm$^2$.
Therefore it is necessary to investigate the timing performances for different beam positions.
We present results for two types of settings.
In one setting we still used a focused beam but shifted it to illuminate a single off-center hole in the collimator mask.    
In the other setting the beam was defocused and simultaneously illuminated several holes of the mask.

\begin{figure}[!ht]\centering
	\includegraphics[width=0.9\columnwidth]{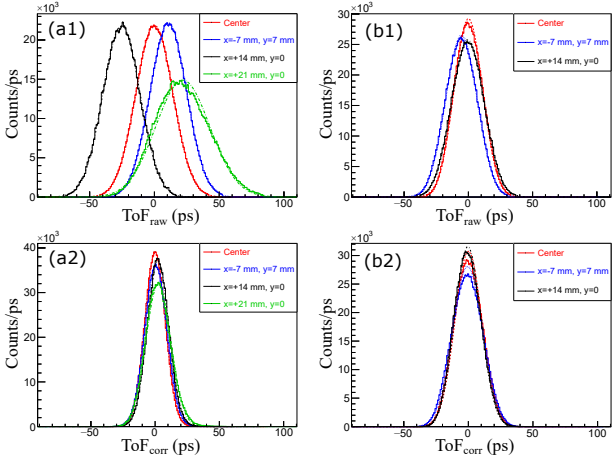}
	\caption{(color online). ToF distributions under different beam conditions using (a) LED and (b) CFD timing methods combined with the same TAC+ADC electronics (1) without and (2) with the time-walk correction. The positions of focused beam are marked in the legends. For better comparing display, all the ToF distributions are normalized to the same total counts but keep their centroids relative to the case of central focused beam for each timing technique. The data corresponds to settings (2--5), and settings (7--9) in Table \ref{tab3}.} \label{TofCompBeam}
\end{figure}

Fig. \ref{TofCompBeam} shows the ToF distributions of central and off-center focused beam hitting positions, both measured with LED and CFD electronic modules.
Without time-walk correction the centroids of the ToF spectra measured with the LEDs shows large variations of $\pm 25$ ps for each position (settings (2--5) in Table \ref{tab3}).
On the other hand, the uncorrected ToF distributions measured with the CFDs are only shifted by $\sim 5$ ps for different positions (settings (7--9) in Table \ref{tab3}).
This shows that the dominant effect in the large spread of data with the LED electronics is due to the time walk.
Any effect due to the position dependence, such as travel time of scintillation photons to the photocathode of PMTs, is minimized by averaging the timing of signals from the four PMTs (see discussion of Eqs. \eqref{ToFTA} and \eqref{ToFTDC}), and is less than 5 ps.

The lower panels of Fig. \ref{TofCompBeam} show the time-of-flight spectra after the time-walk correction.
This was obtained by combining the data of the different beam positions into one data set and then deriving the parameters of the time-walk correction method discussed for Eqs. \eqref{fToF} and \eqref{ToFCorr2}.
Thus, we use the same $ToF_\text{pivot}$ parameter for all settings and can compare the centroids of the corrected ToF distributions.

The time-walk correction significantly reduces the spread of the centroids of measured ToFs, which becomes less than 3 ps for the data taken with LEDs (settings (2--5) in Table \ref{tab3}) and less than 1 ps for the CFD measurements (settings (7--9) in Table \ref{tab3}).
This confirms the conclusion of a small position dependence of the timing beyond any time-walk effect.
As shown in Table \ref{tab3} the detector maintains a good timing resolution even for relatively large position offsets; it is 9.5 ps with LED electronics for a beam position of 2.1 cm off-center.

\begin{figure}[!ht]\centering
	\includegraphics[width=0.9\columnwidth]{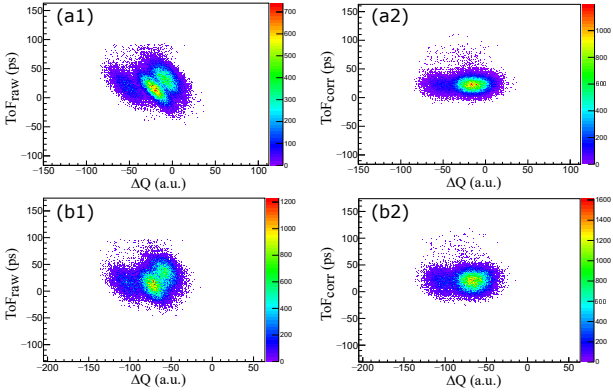}
	\caption{(color online). The correlations between ToF and $\Delta Q$ under the defocused beam condition using (a) LED and (b) CFD (1) without and (2) with the time-walk correction. The defocused beam covers at least four left holes at x=-7 mm, y=$\pm$7 mm and x=-11 mm, y=$\pm$11 mm through the mask. The data corresponds to settings 6 and 10 in Table \ref{tab3}.} \label{TofQDefocus}
\end{figure}

Fig. \ref{TofQDefocus} shows the correlation between the measured ToF and the amplitude difference of PMT signals, $\Delta Q$, for the defocused beam setting when several mask holes were illuminated.
It again illustrates the ability of our correction algorithm to improve the ToF distributions under a beam that illuminates a large area on the scintillator.
The ToF resolution after the time-walk correction, for the spectra including ions going through all illuminated holes, is comparable to that with focused beam for both the measurements with the LED (8.6 ps, setting (6) in Table \ref{tab3}) and CFD (12.2 ps, setting (10) in Table \ref{tab3}) modules.

\subsection{Effects of bias voltages on PMTs}\label{comp5}

In general, the time characteristics of PMTs improve in inverse proportion to the square root of the supply voltage \cite{Matsunaga2017}, which means 100 V modification of the voltage from an initial 1400 V would change the time resolution by $\sim$3.6\% if we assume other conditions are keep constant.
Therefore, it can be expected that the timing performance is not very sensitive to the changes of bias voltage, which was also pointed out by Ref. \cite{Zhao2016}.

In the beginning of the experiment, we adjusted the voltage on each PMT within the range of 1300--1500 V to generate signals with similar amplitude of about 1400 mV, before any cable splitting, for the centered beam setting. 
As shown above, these voltages provide excellent time resolution.
At the end of the beamtime short runs were done changing the bias voltage by $\pm$50 V around those values.
The ToF resolution varied by 0.5 ps, or 4.6\%, for a change of 100 V in bias for the measurement with CFD and TAC+ADC (settings 11 and 12 in Table \ref{tab3}).
This improvement is in agreement with the aforementioned expectation.
In addition, it is observed that the ToF centroids are shifted with varying bias voltage, as expected from the change in the electron transit time in the PMTs.

\subsection{Effects of higher beam rate}\label{comp6}

An ion beam rate between 200 to 600 particles per second was used throughout the test experiment.
In the final stages we increased the beam rates first to 10 kHz, and then to higher than 100 kHz for short runs, to study the timing performance under high rate conditions.
For the 10 kHz beam rate there was no significant change in the performance of the detector (setting (13) in Table \ref{tab3}).
For the beam rate above 100 kHz, the shape of the ToF distributions began to be distorted and the time resolution deteriorated to more than 100 ps (setting (14) in Table \ref{tab3}).
Note that the measurements were taken with a focused beam setting, with all beam ions passing through the scintillator in a small spot of about 2 mm of diameter (dimension of a hole in the collimator mask).

\subsection{Position distribution}\label{posdist}

\begin{figure}[htbp]\centering
	\includegraphics[width=0.9\columnwidth]{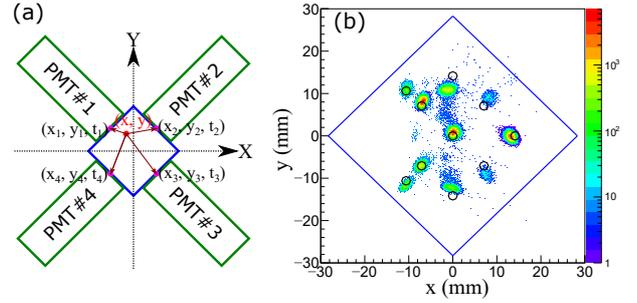}
	\caption{(color online). (a) Illustration of obtaining the hitting position (x and y) on the scintillator from time information (t$_1$, t$_2$, t$_3$ and t$_4$). (b) Distribution of hitting positions on the first scintillator from the LED+TAC+ADC electronics setup and tuning the focused and defocused beam passing through most holes of the front mask shown in Fig. \ref{setup}. The black circles represent the actual holes on mask and the blue square diamond shape represents the plastic scintillator.} \label{position}
\end{figure}

In this section we present result on the capability of determining the beam position on the scintillator from the timing signals of the four coupled PMTs.
As shown schematically in Fig. \ref{position} (a), the relationship between the beam position ($x$, $y$) and the time  ($t_i$) from the emission of light in the scintillator until the photons reach the photocathode of $i^\text{th}$ PMT ($i=1,2,3,4$) is:

\begin{equation} \label{EqPos}
	\left(x-x_i\right)^2+\left(y-y_i\right)^2=v^2t_i^2.
\end{equation}
Making the approximation that light is collected at the center of the photocathode of the PMTs, ($x_i$, $y_i$) is the position for each PMT, with $x_1$=$-A$, $y_1$=$A$; $x_2$=$A$, $y_2$=$A$; $x_3$=$A$, $y_3$=$-A$; $x_4$=$-A$, $y_4$=$-A$ and $A$=14.14 mm. $v$ denotes the speed of light in the plastic. Combining these pieces of information, we can derive:

\begin{align} \label{EqXY}
    x&=\frac{v^2}{4A}\cdot\left(t_1-t_2\right)\left(t_3-t_4\right)\cdot\frac{t_1-t_4+t_2-t_3}{t_1-t_2+t_3-t_4}, \\
	y&=\frac{v^2}{4A}\cdot\left(t_1-t_4\right)\left(t_2-t_3\right)\cdot\frac{t_1-t_2+t_4-t_3}{t_1-t_2+t_3-t_4}.
\end{align}

However, the timing signals measured in our test differ from  $t_i$ because of the different delays in the PMTs and  cables that connect the detectors to the electronics.
Thus, we follow an empirical approach to obtain the beam position using the pattern produced by the mask to calibrate the real positions as the function of variables $x_\text{raw}$ and $y_\text{raw}$,  which are defined as:

\begin{align} \label{EqXYraw}
    x_\text{raw}&=t_1^\text{m}-t_2^\text{m}+t_4^\text{m}-t_3^\text{m}, \\
	y_\text{raw}&=t_4^\text{m}-t_1^\text{m}+t_3^\text{m}-t_2^\text{m}, 
\end{align}
The superscript \emph{m} indicates the measured time signals. The eight TACs with the particular channel assignments used in the electronics setup allow us to measure the time differences between any two PMTs of one detector in the above equations.

The best calibration is obtained with the fitting function of quadratic form in $x_\text{raw}$ and $y_\text{raw}$:

\begin{align} \label{EqXYcal}
    x_\text{cal}=&a_0+a_1x_\text{raw}+a_2y_\text{raw} \nonumber\\
                 &+a_3x_\text{raw}^2+a_4x_\text{raw}y_\text{raw}+a_5y_\text{raw}^2, \\
	y_\text{cal}=&b_0+b_1x_\text{raw}+b_2y_\text{raw} \nonumber\\
	             &+b_3x_\text{raw}^2+b_4x_\text{raw}y_\text{raw}+b_5y_\text{raw}^2.
\end{align}
Data from the focused and defocused beams settings, with LED+TAC+ADC electronics, are combined to obtain enough  calibration points.

Fig. \ref{position} (b) shows the final distribution of the calibrated positions.
The results give very good position resolution of about 0.7 mm ($\sigma$) for most spots in Fig. \ref{position} (b).

It is also clear that the calibration does not accurately reproduce the holes' pattern on the mask.
In addition to limitations in the calibrating algorithm used in Eqs. \eqref{EqXYraw}--\eqref{EqXYcal}, another possible source of error is the time-walk effect of the LED modules that could be reduced by using the measured signal amplitudes.

\section{Summary}

In this work, we presented the results of tests from a new timing detector system developed for ToF-$B\rho$ mass measurement experiments.
Each timing detector consists of a thin square organic scintillator and four fast photomultiplier tubes (PMTs) coupled to its sides.
A systematic test of the detector's performance was conducted with a $^{48}$Ca beam with energy of 90 MeV/u at the NSCL.

The best time resolution ($\sigma$) of 7.5 ps was achieved using an electronics setup consisting of a leading-edge discriminator (LED), and the combination of a time-to-amplitude converter and a amplitude-to-digital converter (TAC+ADC) modules for digitization.
Achieving this resolution required a correction to the time-walk effect using the signal amplitude recorded for each PMT with a charge-to-digital converter (QDC) module, and a newly developed correction method.
The results represent a significant improvement from the time resolution of $\sim$30 ps of the detectors used in previous experiments \cite{Matos2012}.

The detector system was also tested using constant-fraction discriminator (CFD) modules.
These showed a robust response to time-walk effect, and provided timing with a resolution of $\sim$11 ps without any correction for PMT signal amplitude.
Such systems represent a good alternative for application that do not require timing resolution below 10 ps.

We evaluated the position dependence of the timing response by illuminating different spots across the surface area of the scintillator.
The detector showed a relatively small decrease of its resolution ($\leqslant 25\%$) for positions more than 2 cm off-center.
These settings provide a good approximation to experiments with secondary beams when the beam spot with a radius of $\sim$1 cm, so we expect that the ToF resolution of the detector in the mass measurement experiment will be $\sim$12 ps. The performance of the new detector system represents a significant improvement of the time resolution of the detectors used in ToF-$B\rho$ mass measurements experiments at the NSCL, and an important step to achieve the goal of a mass resolution of $1.0\times 10^{-4}$.

In addition, we also attempted to derive the beam hitting position from the measured time information of the four PMTs coupled to one scintillator.
We obtained a good position resolution ($\sigma$) of about 0.7 mm, albeit with systematic shifts in the absolute position measurement that is the focus of ongoing work. 

\section*{ACKNOWLEDGEMENTS}

We are grateful to the operation staff of the NSCL for providing the $^{48}$Ca beam.
This work was supported in part by the US Department of Energy under Grant No. DE-SC0020406, and the US National Science Foundation under Grant Nos. PHY-1712832, PHY-1714153 and PHY-1430152.
K. Wang acknowledges the support by FRIB-CSC Research Fellow Program.

\section*{References}

\bibliography{mybibfile}

\end{document}